\documentclass[3p, twocolumn]{elsarticle}

\usepackage{amssymb}
\usepackage{bm}

\makeatletter
\def\ps@pprintTitle{%
 \let\@oddhead\@empty
 \let\@evenhead\@empty
 \def\@oddfoot{}%
 \let\@evenfoot\@oddfoot}
\makeatother

\date{October 23, 2014}

\newcommand{\gs}{86mm}
\newcommand{\etal}{\textit{et al.}}

\begin{document}

\begin{frontmatter}

\title{Critical analysis of the slope method for estimation of ice-water interfacial energy from ice nucleation experimental data (with reviews)} 



\author[mymainaddress,mysecondaryaddress]{Tom\'a\v{s} N\v{e}mec\corref{mycorrespondingauthor}}
\cortext[mycorrespondingauthor]{Corresponding author}
\ead{nemec@it.cas.cz}

\address[mymainaddress]{Institute of Thermomechanics of the CAS, v. v. i., Dolej\v{s}kova 5, 18200 Praha 8, Czech Republic}
\address[mysecondaryaddress]{New Technologies - Research Centre, University of West Bohemia, Univerzitn\'i 8, 30614 Plze\v{n}, Czech Republic}

\begin{abstract}
An established procedure for the estimation of ice-water interfacial energy based on evaluation of the slope of the experimental ice nucleation rate data versus scaled temperature is critically analyzed in this work. An inconsistent estimate of the ice-water interfacial energy is found in the work of Murray \etal~[Phys. Chem. Chem. Phys., 2010, 12, 10380--10387]. The source of the inconsistency is identified in an inappropriate regression method used for experimental ice nucleation data fitting, a correct estimate of the ice-water interfacial energy is presented, and limits of the slope method are discussed. 

\end{abstract}


\end{frontmatter}


\section{Introduction}

The classical nucleation theory (CNT) provides a theoretical link between the ice nucleation rates and the thermophysical properties of the ice-water system, i.e. the density of the solid phase, the diffusion coefficient of the liquid phase, the equilibrium pressure, and the interfacial energy \cite{ickes15}. Therefore, in cases where experimental measurements of the nucleation rate are available, CNT presents an indirect, theoretical approach for the estimation of the interfacial energy provided that the other above mentioned thermophysical properties are known.

For ice and water, the interfacial energy measurements reported in literature are restricted to the triple point temperature \cite{granasy02}. The estimation of the interfacial energy for lower temperatures in the supercooled region, was possible only due to ice nucleation rate measurements \cite{murray10} or molecular simulations of ice nuclei formation \cite{li11, espinoza16}.

In this work, a method for the estimation of ice-water interfacial energy from the experimental nucleation rates \cite{pant06, parsons06} will be investigated that is based on the evaluation of the slope of the logarithm of the measured nucleation rate $\ln J$ vs. the function $T^{-3}(\ln(S))^{-2}$ of the experimental temperature $T$, where $S$ is the ratio of water and ice saturation pressures. It will be referred to as the \textit{slope method} in the following text. Several authors estimated the ice-water interfacial energies based on the slope method \cite{murray10, murray12, manka12, bhabhe13} covering the temperature range 200 -- 240~K. I will study the validity of the ice-water interfacial energy estimates by the slope method in this paper. Particularly, I intend to show that the ice-water interfacial energy estimate of Murray \etal~\cite{murray10} suffers from internal inconsistency and needs to be reconsidered. The nature of this inconsistency will be studied in detail to point out the weak spots of the slope method.


The paper is arranged in the following way. First, the equations of the slope method for interfacial energy estimation are derived in section \ref{se:method}. In section \ref{se:proof}, a proof is presented showing the inconsistency of the interfacial energy estimate by Murray \etal~\cite{murray10}. In section \ref{se:discuss}, the source of the inconsistency is identified, a correct estimate of the ice-water interfacial energy is presented, and the limits of applicability of the slope method are discussed. 

\section{Slope method} \label{se:method}

The slope method \cite{pant06, parsons06} for the estimation of the ice-water interfacial energy is based on the CNT nucleation rate equation \cite{kashchiev00}
\begin{equation} \label{eq:nr}
J = J_0 \exp \left( -\frac{W^{\star}}{kT} \right)
\end{equation}
relating the nucleation rate $J$ [m$^{-3}$s$^{-1}$] to the nucleation work of the critical cluster $W^{\star}$ [J]. The pre-factor $J_0$ [m$^{-3}$s$^{-1}$] reflects the kinetics of the cluster growth and $T$ [K] is absolute temperature. By using the CNT expression for the critical nucleation work $W^{\star} = \frac{16 \pi \gamma^3 v^2}{3 (kT \ln S)^2}$, Eq. (\ref{eq:nr}) can be rearranged to
\begin{equation} \label{eq:nrln}
\ln J = \ln J_0 - \frac{16 \pi \gamma^3 v^2}{3 k^3 T^3 (\ln S)^2}
\end{equation}
where $\gamma$ [J/m$^2$] is the ice-water interfacial energy, $v$ [m$^3$] is the molecular volume of the solid phase, and the ratio of saturation pressures $S = p^{eq}_{l}/p^{eq}_{s}$ represents the supersaturation of the liquid. Here, the vapor-liquid equilibrium pressure is denoted $p^{eq}_{l}$, and the vapor-solid equilibrium pressure is denoted $p^{eq}_{s}$. The vapor-solid equilibrium pressure of cubic ice \cite{shilling06} is used in this work to retain consistency with the work of Murray \etal~\cite{murray10}. However, the structure of the clusters in ice nucleation is still under debate \cite{moore11}.

To transform Eq.~(\ref{eq:nrln}) into the relations of the slope method, the following assumptions are made. The interfacial energy is taken constant, i.e. independent of temperature. The density of the solid phase is taken constant. And the pre-factor is taken constant. Since the size of the temperature interval of the experimental nucleation data analyzed by the slope method is typically a few Kelvins, the assumptions of the constancy of the thermophysical properties over this narrow temperature interval are plausible, and Eq.~(\ref{eq:nrln}) can be written in the form
\begin{equation} \label{eq:nrln3}
\ln J = n + m t_s
\end{equation}
where the parameters
\begin{equation} \label{eq:n}
n = \ln J_0
\end{equation}
and
\begin{equation} \label{eq:m}
m = -\frac{16 \pi \gamma^3 v^2}{3k^3}
\end{equation}
are both due to the above assumptions constant, and
\begin{equation} \label{eq:ts}
t_s = \frac{1}{T^3 (\ln S)^2}
\end{equation}
is a scaled temperature. As a result, the slope method is simply a fit of the experimental ice nucleation rate data to the linear function~(\ref{eq:nrln3}) in $t_s$, giving the parameters $n$ and $m$. The ice-water interfacial energy $\gamma$ corresponding to the analyzed ice nucleation data is recovered from the slope $m$ according to Eq.~(\ref{eq:m}), and the pre-factor $J_0$ is given by the absolute value $n$ according to Eq.~(\ref{eq:n}).

\section{Inconsistency of interfacial energy estimates} \label{se:proof}

The validity of the slope method estimates of the ice-water interfacial energy based on the experimental nucleation data as published by Murray \etal~\cite{murray10} will be investigated in this section.

\begin{figure}[t]
	\centering
	\includegraphics[width=\gs]{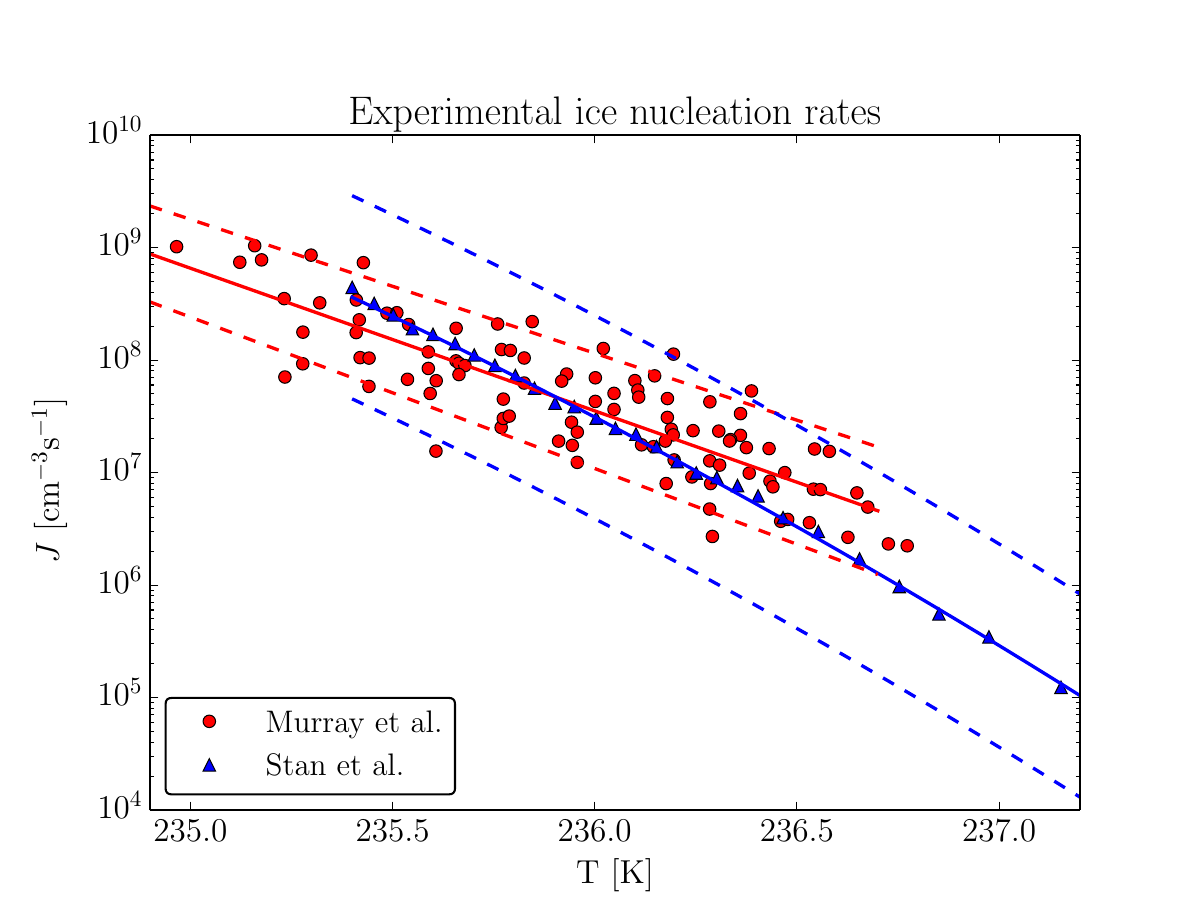}
	\caption{Experimental ice nucleation rates reported by Murray \etal~\cite{murray10} and Stan \etal~\cite{stan09}. The full lines show fits of the experimental data, and the dashed lines delimit the standard deviation bands. The fit of Murray \etal~\cite{murray10} is used as presented in their Fig.~4. The data of Stan \etal~\cite{stan09} as presented in their Fig.~8 are shown with the reported $\pm$0.4~$^{\circ}$C standard deviation.
	}
	\label{fig:mse}
\end{figure}

I have identified inconsistent estimates of interfacial energy for two specific sets of ice nucleation data investigated by Murray \etal~\cite{murray10}, i.e. their own experimental data and the data measured by Stan \etal~\cite{stan09}, respectively. The two experimental data sets are plotted in Fig.~\ref{fig:mse}. By using the slope method, Murray \etal~\cite{murray10} estimated ice-water interfacial energy $\gamma_M$ = 20.8 $\pm$ 1.2 mJ/m$^2$ from their own experimental data, and ice-water interfacial energy $\gamma_S$ = 23.7 $\pm$ 1.1 mJ/m$^2$ from the experimental data of Stan \etal~\cite{stan09}, respectively. It is the gap between the two intervals, $\gamma_M$ and $\gamma_S$, that renders the two interfacial energy estimates inconsistent. In other words, as I will show below in detail, since the initial two experimental datasets used as input for the evaluation of the respective interfacial energies do overlap to a large extent, which is seen in Fig. \ref{fig:mse}, it is impossible to calculate two interfacial energy estimates by the slope method that form two disjunct intervals, as is the case with $\gamma_M$ and $\gamma_S$.


To prove the particular inconsistency in interfacial energy estimates $\gamma_M$ and $\gamma_S$ suggested above, I will frame general requirements for a physically-relevant deduction of the ice-water interfacial energy from the nucleation rate according to the CNT. Mathematically, the deductions will utilize elementary properties of continuous, strictly monotonic functions in a reductio ad absurdum type of logical proof. The slope method analysis will present a limiting case of this general scenario.

Let us denote $g$ a continuous, strictly monotonic function of temperature $g$: $T$ $\rightarrow$ $\gamma$, which stands for the functional dependence of the interfacial energy on temperature. Similarly, let $j$ be a continuous, strictly monotonic function of temperature $j$: $T$ $\rightarrow$ $J$, which stands for the functional dependence of the nucleation rate on temperature, e.g. in the form of Eq.~(\ref{eq:nr}). Let the temperature interval [$T_1$, $T_2$] be a subdomain of the two functions $g$ and $j$ that corresponds to the temperature range of the analyzed ice nucleation experiment. Both functions $g$ and $j$ are bijective by their above definitions. In other words, they present a one-to-one correspondence between the elements of subdomain [$T_1$, $T_2$] and the elements of functional images $g([T_1, T_2])$ and $j([T_1, T_2])$, respectively. Therefore, inverse function $j^{-1}$: $J$ $\rightarrow$ $T$ exists and it is bijective as well. Furthermore, composite function $g_j$ = $g(j^{-1})$: $J$ $\rightarrow$ $\gamma$ is bijective. Here, $g_j$ is the formal representation of a general theoretical method to deduce the interfacial energy from the nucleation rate in temperature range [$T_1$, $T_2$].

Now, let $\bm{J_1}$ = [$J_{11}$, $J_{12}$] and $\bm{J_2}$ = [$J_{21}$, $J_{22}$] be two intervals of nucleation rates that possess a non-empty intersection $\bm{J_0}$ = $\bm{J_1} \cap \bm{J_2} $ $\neq \emptyset$. Let $\bm{\gamma_1}$ = [$\gamma_{11}$, $\gamma_{12}$] and $\bm{\gamma_2}$ = [$\gamma_{21}$, $\gamma_{22}$] be $g_j$'s functional images of intervals $\bm{J_1}$ and $\bm{J_2}$, i.e. $\bm{\gamma_1}$ = $g_j(\bm{J_1})$ and $\bm{\gamma_2}$ = $g_j(\bm{J_2})$, respectively. Intervals $\bm{\gamma_1}$ and $\bm{\gamma_2}$ represent the ranges of interfacial energies deduced from the two nucleation rate data sets $\bm{J_1}$ and $\bm{J_2}$ according to a general method $g_j$. Then $\forall J \in \bm{J_0}$ it holds that $g_j(J) \in \bm{\gamma_1}$ and $g_j(J) \in \bm{\gamma_2}$. In other words, for any value of the nucleation rate $J$, which belongs to the intersection of intervals $\bm{J_1}$ and $\bm{J_2}$, the deduced value of interfacial energy according to method $g_j$ belongs to interval $\bm{\gamma_1}$ and also to interval $\bm{\gamma_2}$. Therefore, since element $\gamma = g_j(J)$ belongs to both intervals $\bm{\gamma_1}$ and $\bm{\gamma_2}$, intersection $\bm{\gamma_0}$ = $\bm{\gamma_1} \cap \bm{\gamma_2} $ has at least one element $\gamma$ and, therefore, $\bm{\gamma_0}$ must be non-empty, i.e. $\bm{\gamma_0} \neq \emptyset$. 

For the particular case of the ice nucleation experimental data of Murray~\etal~\cite{murray10} and Stan~\etal~\cite{stan09} the intersection of the measured nucleation rate data sets is not empty, as can be seen in Fig.~\ref{fig:mse}; the measured values of the nucleation rate in the two datasets clearly overlap to a large extent. According to the above reasoning the range of interfacial energies deduced from the Murray~\etal~\cite{murray10} nucleation data must have an non-empty intersection with the range of interfacial energies deduced from the Stan~\etal~\cite{stan09} nucleation data provided that the method used to deduce the interfacial energy assures that the temperature dependency of the interfacial energy and the nucleation rate is strictly monotonic. The strict monotonicity of both the nucleation rate and the interfacial energy 
is a natural requirement conforming to the physical reality; the nucleation rate decreases with increasing temperature and the interfacial energy increases with increasing temperature \cite{ickes15}.

However, the slope method assumes, rather unphysically, that the interfacial energy is constant, and not strictly monotonic as supposed in the general case above. Under such assumption, the considerations presented above result into an even stronger requirement on the deduced interfacial energies from two intersecting nucleation rate datasets. Any constant function can be viewed as a limit of a sequence of strictly monotonic functions, e.g. a sequence of linear functions $f_n = C + a_n (T - T_0)$ with slope $a_n$ decreasing as $1/n$ for $n \rightarrow \infty$. For every element of the sequence the above reasoning applies and the deduced interfacial energy ranges must possess a non-empty intersection. In the limit of the constant function the intersection shrinks to a single value of interfacial energy, common to both datasets. As a result, the slope method must produce interfacial energies equal to each other within their standard deviation ranges from two overlapping datasets of nucleation rates. However, we already stated that the estimates of the interfacial energy of the slope method, $\gamma_M$ and $\gamma_S$, as published by Murray \etal~\cite{murray10} are not equal within their standard deviations, which is in contradiction with the outcome of the above reasoning. Therefore, the slope method procedure as preformed by Murray \etal~\cite{murray10} must contain a hidden inconsistency, an erroneous step that introduces additional uncertainty. Its nature will be discussed in the following section.



\section{Discussion} \label{se:discuss}

The fit of the experimental ice nucleation-rate data to the linear function (\ref{eq:nrln3}) forms the basis of the slope method. The fit results in two parameters $n$ and $m$ that will be discussed in this section. First, the evaluation of the slope $m$ resulting in the estimation of ice-water interfacial energy (\ref{eq:m}) will be investigated with the goal of resolving the inconsistency found in Sec.~\ref{se:proof}. Second, the evaluation of the nucleation rate pre-factor from the term $n$ according to Eq.~(\ref{eq:n}) will be analyzed.

\subsection{Interfacial energy} \label{se:gamma}
Although Murray \etal~\cite{murray10} do not specify the particular regression method that they used to perform the linear fit to their data yielding their reported slope $m_{OLS} = -$ (6.02 $\pm$ 0.36)$\times$10$^7$~K$^3$, their result can be reproduced with the widely used ordinary least squares (OLS) method. By using the OLS method implemented in Python (package \texttt{scipy.stats.linregress}) to fit the Murray \etal~\cite{murray10} experimental data to the function (\ref{eq:nrln3}) we can check that the above mentioned value of $m_{OLS}$ is recovered. Therefore, let us assume that the OLS method was indeed used by Murray \etal~\cite{murray10} in their analysis.

\begin{figure}[t]
	\centering
	\includegraphics[width=\gs]{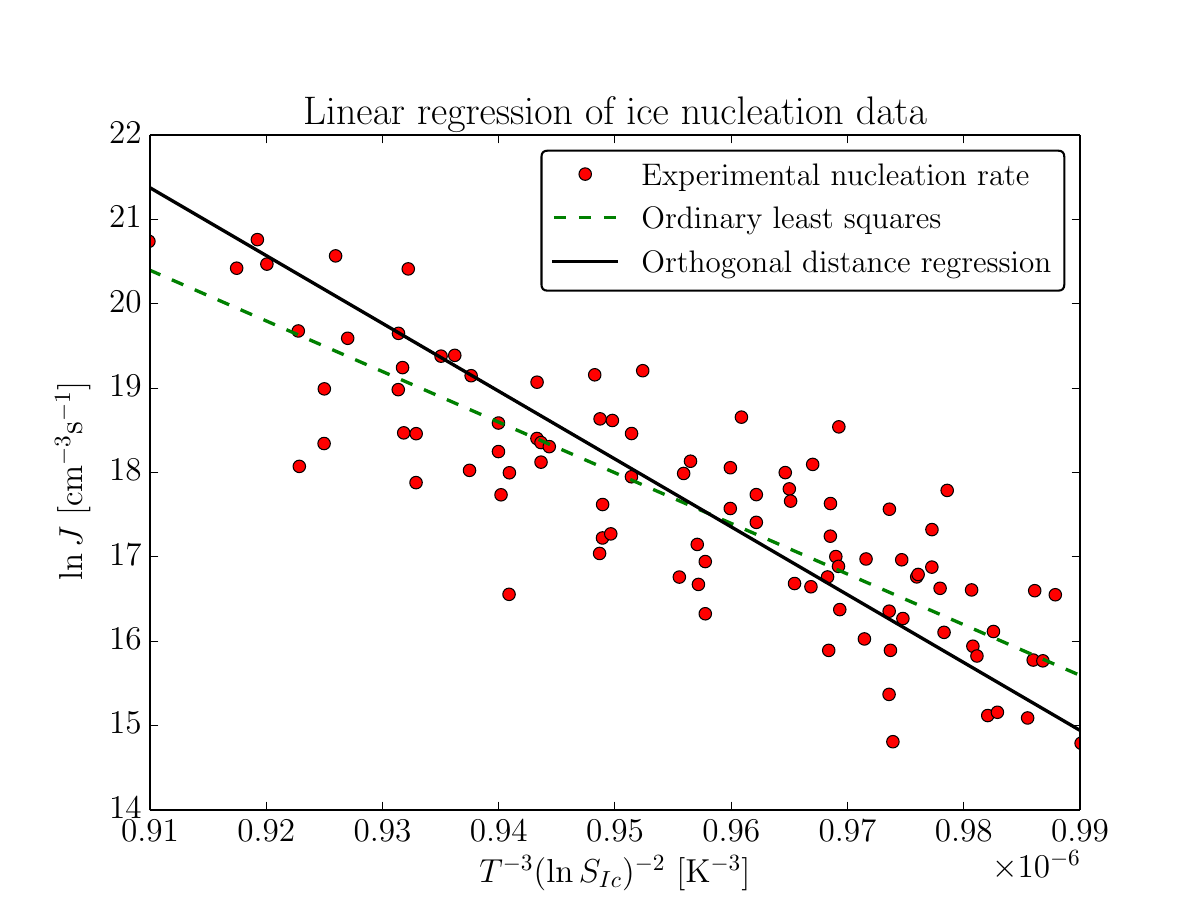}
	\caption{Comparison of the ordinary least squares regression method (dashed line) and the orthogonal distance regression method (full line) for the estimation of the slope of the ice nucleation data of Murray \etal~\cite{murray10}.
	}
	\label{fig:munc}
\end{figure}

The OLS method was derived with the assumption that the observation of the independent variable, i.e. the temperature in this case, is error-free. From this point of view the applicability of the OLS method is highly questionable for nucleation-rate experimental data, because the error in the temperature measurement of an ice nucleation experiment is typically larger than $\pm$0.4~$^{\circ}$C, and it is identified as the main source of uncertainty in the nucleation measurement \cite{stan09}.

The regression method derived with the assumption of non-zero observational errors in the independent variable is the orthogonal distance regression method (ODR), also known as errors-in-variables modeling, or total least squares \cite{huffel91}. The Python ODR implementation (\texttt{scipy.odr.odrpack}) applied to Murray \etal~\cite{murray10} ice nucleation data results in the slope $m_{ODR}=-$ (8.04 $\pm$ 0.48)$\times$10$^7$~K$^3$. 

The difference in the slope estimates between the OLS and ODR methods for the Murray \etal~\cite{murray10} ice nucleation data is shown in Fig.~\ref{fig:munc}. The ice-water interfacial energy corresponding to the $m_{ODR}$ is $\gamma_{ODR}$ = 22.9 $\pm$ 0.5 mJ/m$^2$ according to the slope method, Eq. (\ref{eq:m}). And after including the uncertainty in the cubic ice sublimation pressure \cite{shilling06} ($\pm$ 0.8 mJ/m$^2$), the final estimate of ice-water interfacial energy using the ODR method is $\gamma_{ODR}$ = 22.9 $\pm$ 1.3 mJ/m$^2$. This new ODR estimate is roughly 10 \% higher that the OLS estimate $\gamma_{OLS}$ = 20.8 $\pm$ 1.2 mJ/m$^2$ calculated from $m_{OLS}$. The new value $\gamma_{ODR}$ is sufficiently close to the value $\gamma_S$ Murray \etal~\cite{murray10} estimated from the nucleation data of Stan \etal~\cite{stan09}, and therefore $\gamma_{ODR}$ does not suffer from the contradiction described in Sec.~\ref{se:proof}.

Note that the Stan data~\cite{stan09} are fitted correctly even with the OLS method that fails for fitting the Murray data \cite{murray10}. 
We can check that the OLS method gives the slope $m = - 8.952\times$10$^7$~K$^3$ and the ODR method results in $m = - 8.967 \times$10$^7$~K$^3$, which both correspond to the interfacial energy $\gamma_S$  = 23.7 mJ/m$^2$ reported by Murray \etal~\cite{murray10} for the nucleation data of Stan \etal~\cite{stan09}. 
The Stan data, as presented in Fig.~\ref{fig:mse} (and in their Fig.~8 \cite{stan09}), on the contrary to the raw data by Murray \etal~\cite{murray10}, are already averaged from a set of more than 37 thousand freezing experiments, effectively removing the large uncertainty in experimental temperature measurement. The Stan data are therefore suitable for fitting even by the OLS method.

\subsection{Nucleation rate pre-factor} \label{se:j0}

\begin{figure}[t]
	\centering
	\includegraphics[width=\gs]{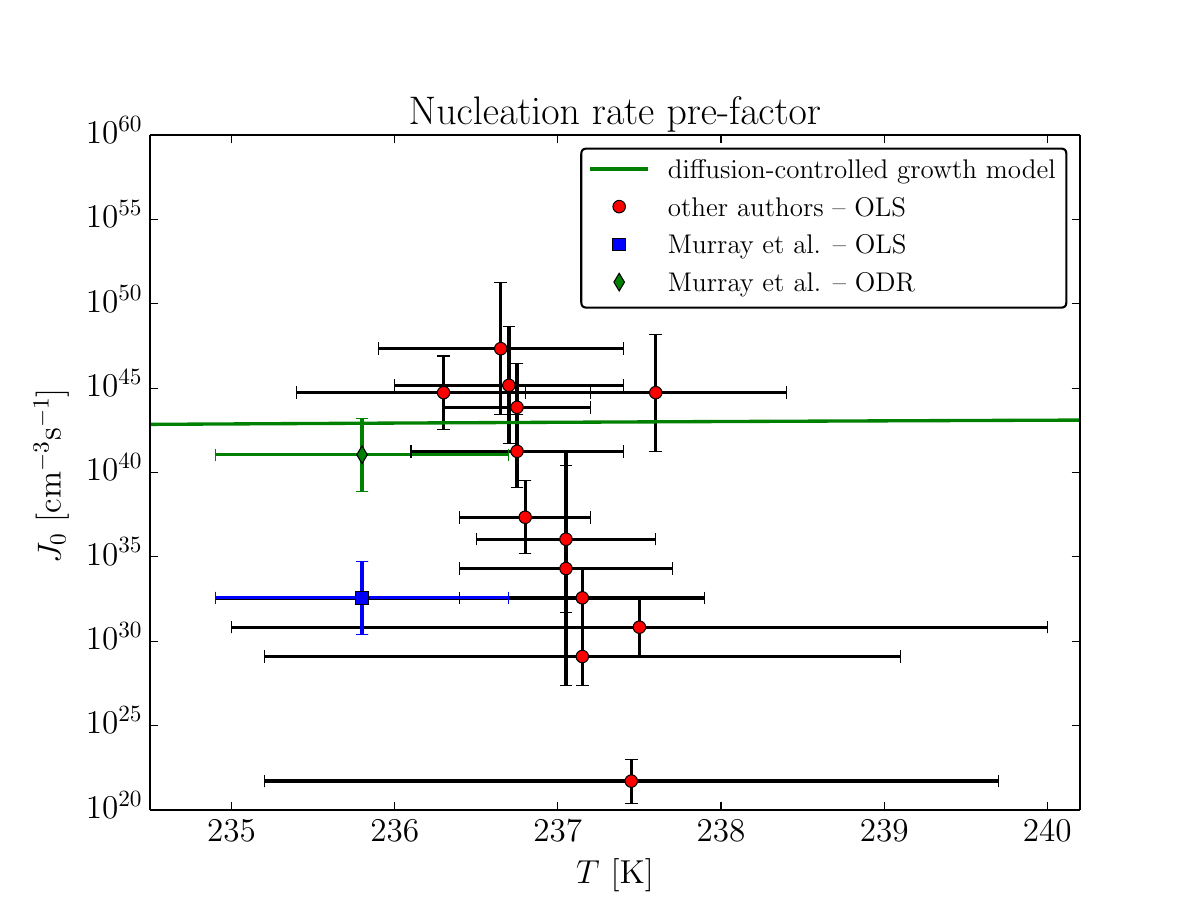}
	\caption{Pre-factor $J_0$ estimated by Murray~\etal~\cite{murray10} with the slope method (as given in their Table~2) from the ice nucleation data available in literature (red circles). Horizontal errorbars show the temperature range of the input experimental nucleation-rate datasets. The vertical errorbars denote the uncertainty of $J_0$ fit. The green full line shows the pre-factor according to the volume-diffusion controlled growth model \cite{kashchiev00}. The blue square shows the OLS-based fit of the pre-factor, and the the green diamond shows the ODR-based fit of the pre-factor from the experimental data of Murray~\etal~\cite{murray10}.
	}
	\label{fig:mpf}
\end{figure}

The slope method concurrently fits two parameters, $n$ and $m$, from a given data set of experimental nucleation rates. The two parameters are therefore coupled and we can expect that the larger an error in the estimation of $m$ we make, the larger the corresponding error in $n$ we get. It was shown in the preceding section that the interfacial energy estimate $\gamma_M$ suffered from an error due to the inappropriate statistical processing of the experimental data. Therefore, an error was introduced in the corresponding estimate of nucleation rate pre-factor $J_0$ as well.

The difference between the pre-factor estimates using the OLS and the ODR regression methods is shown in Fig.~\ref{fig:mpf} for the experimental data of Murray \etal~\cite{murray10}. The fitted parameters are $n_{OLS}$ = 75 $\pm$ 5, and $n_{ODR}$ = 94.5 $\pm$ 5, respectively. By evaluating the pre-factor according to Eq.~(\ref{eq:n}), we find that the OLS-based estimate of $J_0$ is roughly 9 orders of magnitude lower than the ODR-based result. It is the ODR-based estimate of $J_0$ that is within its uncertainty equal to an independent model of $J_0$, i.e. the volume-diffusion based description of the cluster growth \cite{kashchiev00} (p.~141), as shown in Fig.~\ref{fig:mpf}.

\subsection{Final remarks} \label{se:fr}

Not only the already discussed regression issue is the source of uncertainty in the slope method. The assumptions of the method itself, as summarized in Sec.~\ref{se:method}, play a role as well. The constancy of the physical properties assumed by the slope method is plausible for small temperature ranges of the analyzed experimental nucleation rates. As the temperature range of the experiment increases, the interfacial energy, ice density, and the pre-factor are getting less accurately approximated by a constant. It is therefore desirable for the slope method to analyze temperature ranges as small as possible. On the other hand, to evaluate the slope of the ice nucleation data from the cloud of scattered experimental nucleation rates one needs as large temperature ranges as possible, because the uncertainty of the fit of the slope gets larger as the temperature range of the analyzed data decreases. Obviously, both requirements, i.e. having the temperature interval small for the assumptions of the slope method to be valid and having the temperature interval large for a precise evaluation of the slope, cannot be simultaneously satisfied. And uncertainty due to this fact will be inevitably present in the results of the slope method.

Fig.~\ref{fig:mpf} shows the $J_0$ estimates from all the ice nucleation-rate datasets in the temperature range 235 -- 240~K available in literature as calculated by Murray \etal~\cite{murray10}. A huge scatter in the pre-factor estimates spanning roughly 25 orders of magnitude can be observed. However, the pre-factor is only slightly temperature dependent in the case of nucleation in liquids \cite{kashchiev00} (p.~199). According to the above-mentioned diffusion-based growth model the pre-factor depends linearly on the self-diffusion coefficient of supercooled water, which was measured down to 238~K with accuracy less than 1\% \cite{price99}. Therefore, the volume-diffusion based growth model gives us an estimate of the pre-factor $J_0$ with negligible uncertainty compared to the uncertainty of 25 orders of magnitude reported in Murray \etal ~\cite{murray10}. Obviously, the scatter of $J_0$ estimates as derived by Murray \etal~\cite{murray10} and shown in Fig.~\ref{fig:mpf} is non-physical, and reflects the errors in implementing the slope method.

Finally, the slope method was also used by Manka \etal~\cite{manka12} to estimate the ice-water energy of 15.6 mJ/m$^2$ from their ice nucleation data measured in the temperature range 202 -- 215~K. The authors do not report any uncertainty of their interfacial energy estimate; they just state "we analyzed all of our data using the same formalism" as Murray \etal~\cite{murray10}. Therefore, a similar regression issue as in the above-discussed case of Murray \etal~\cite{murray10} interfacial energy estimate is likely to arise. Also, the relatively large temperature interval of the Manka data collides with the assumptions of constancy of the thermophysical properties of the slope method. But most importantly, the CNT formulation used in the slope method as presented in Sec.~\ref{se:method} is not valid for ice nucleation data measured at such low temperatures due to the omitted pressure effect, as shown recently \cite{nemec13}. The pressure-related terms in the CNT formulation present a considerable contribution to the nucleation work and, therefore, account for a shift in the CNT-predicted ice nucleation rate at low temperatures. The errors in CNT formulation are then projected to errors in the ice-water interfacial energy estimates of the slope method.

\section{Conclusion}

A method for the estimation of the ice-water interfacial energy based on the classical nucleation theory and on the evaluation of the slope of the experimental ice nucleation rates vs. a scaled temperature was thoroughly analyzed in this work. A logical contradiction in the interfacial energy estimates was identified in the work of Murray \etal~\cite{murray10} that is related to the linear regression algorithm utilized for the evaluation of the slope of experimental ice nucleation rate data. The contradiction was removed by using the orthogonal distance regression method for a proper evaluation of the slope instead of the ordinary least squares method. The corrected estimate of the interfacial energy is by 10 \% (2.1 mJ/m$^2$) higher than the original value reported by Murray \etal~\cite{murray10}.

In the light of the findings of this work, estimation methods \cite{huang95, nemec13} utilizing the absolute values of the experimental nucleation rates instead of the temperature derivative, and using a theoretical model of the nucleation rate pre-factor instead of its fit from experimental nucleation data, present a safer way of deducing the interfacial energy from nucleation rate data avoiding uncertainties inherently contained in the slope method.

\section*{Acknowledgement}

The author would like to acknowledge the institutional support RVO:61388998 of the Institute of Thermomechanics, v.v.i., support by the Czech Science Foundation (project GAP101/10/1819), by the CENTEM project (reg. no. CZ.1.05/2.1.00/03.0088) co-funded by the ERDF as part of the Ministry of Education, Youth and Sports OP RDI programme and, in the follow-up sustainability stage, supported through CENTEM PLUS (LO1402) by financial means from the Ministry of Education, Youth and Sports under the National Sustainability Programme I, and by the POLYMEM project (reg. no. CZ.1.07/2.3.00/20.0107) co-funded by the ESF as part of the Ministry of Education, Youth and Sports ECOP programme.


\appendix

\section{Supporting material}

The history of the review process of the manuscript of this critical analysis is summarized below. It is quite interesting to observe how the original endeavor to discover the scientific truth changed more to a social experiment over time.

The summary of submissions, referees' comments, editors' decisions and author's appeals presented in the following sections has a certain scientific value of its own. The manuscript was submitted to three different journals between 2013 and 2017 and received seven reviews. All reviews were strongly negative, highly correlated, and the publication in all three journals was rejected by the respective editors. This way, the anonymous referees managed to defend the results of the slope method by Murray \etal \cite{murray10}. However, the same group of authors published new measurements of ice nucleation rate in their 2016 paper [Atkinson, J. D., Murray, B. J., and O'Sullivan, D.: Rate of Homogenous Nucleation of Ice in Supercooled Water, The Journal of Physical Chemistry A, 120, 6513-6520, 2016], with rather surprising results. On the contrary to what all the seven referees successfully defended, the slope of the new 2016 ice nucleation data coincides perfectly with the corrected slope of ice nucleation rates deduced in this critical analysis (Fig.~\ref{fig:munc}) by reanalyzing the original 2010 ice nucleation measurements and applying the ODR regression.

\section{History of the review process}

The original, 2 pages long version of the manuscript was submitted to Physical Chemistry Chemical Physics (PCCP) as a comment to the Murray's paper \cite{murray10}. The manuscript was under review from May 2013 to January 2014 and received 5 reviews. The editor rejected the publication of the manuscript and advised the author to \emph{publish this comment to a specialized journal in the field}.

An extended, 6 pages long manuscript was submitted to Journal of Crystal Growth and it was under review between October 2014 and March 2015. The publication was rejected and both the reviewer and the editor expressed the opinion that \emph{this paper should be submitted to PCCP journal, because the data previously published in PCCP were reanalyzed in this paper}.

Finally, the manuscript was submitted to Advances in Physical Chemistry in July 2016. After receiving one review, the manuscript was rejected in June 2017.

The considerable amount of work performed by the referees to defend the results of Murray \etal \cite{murray10} is presented in \ref{ap:pccp} -- \ref{ap:apc}. The referees judged the findings presented in this critical analysis as wrong, incorrect, extremely naive, inappropriate, not relevant. Also, the referees mention that the paper takes a narrow view of the subject (\ref{ap:pccpr4}), it reports no important new physical insight (\ref{ap:pccpr3}), it fails to address the much larger problem of systematic error (\ref{ap:pccpr5}). Moreover, the referees state that the paper does not reach the standard of a scientific paper, or even a technical comment (\ref{ap:jcgr1}), it is misleading in the claim that there is a problem (\ref{ap:jcgr1}) and the 'proof' in section 3 which is presented should be deleted in its entirety (\ref{ap:apcr1}). Finally, the referees react to the writing style as aggressive, inflammatory, arrogant, and one referee feels that \emph{the author seems to have taken the stance that much of the rest of the community are idiots} (\ref{ap:jcgr1}).

On July 13, 2016, Murray's group published an article [Atkinson, J. D., Murray, B. J., and O'Sullivan, D.: Rate of Homogenous Nucleation of Ice in Supercooled Water, The Journal of Physical Chemistry A, 120, 6513-6520, 2016] that shows new measurements of the ice nucleation rate in the same temperature range as in their 2010 paper. The cold stage experiment used already in their 2010 paper was upgraded. The authors stress in the introduction that they \emph{paid special attention to reducing and quantifying the uncertainty in both the rate and the temperature of nucleation}. The results are quite surprising; the temperature dependence of the new 2016 nucleation rate data is the same as the temperature dependence deduced by the ODR method in this critical analysis (Fig.~\ref{fig:munc}) from the 2010 data.

The authors do actually discuss the difference in slope between their 2010 and 2016 results stating that the 2010 results \emph{are consistent with the new data but the temperature dependence is not as strong}. They claim that \emph{the explanation for this is that Murray \etal \cite{murray10} used a different method of determining average droplet volume}. The authors make no attempt to deduce the ice-water interfacial energy from their new ice nucleation experiments this time.


\section{Physical Chemistry Chemical Physics (May 2013 -- January 2014)} \label{ap:pccp}
This section concerns the first version of this paper, which was written as a comment to the Murray 2010 paper \cite{murray10}. Although the comment was limited to two pages in length, it already included the main ideas, i.e. the contradiction in deriving two distinct results from two almost identical datasets, and the explanation for OLS vs. ODR regression method.

\subsection{Manuscript submission -- May 21, 2013}
Dear PCCP Editors,

I would like to submit the paper Comment on "Kinetics of the homogeneous freezing of water"
[Phys. Chem. Chem. Phys., 2010, 12, 10380--10387] by Tom\'a\v{s} N\v{e}mec for your consideration for publication in Physical Chemistry Chemical Physics as a comment.

I found certain inconsistencies regarding the uncertainty in the ice-water
interfacial tension estimation in the work of Murray et al. [Phys. Chem. Chem.
Phys., 2010, 12, 10380--10387] that I am addressing in this comment. I show that
the method of estimating the interfacial energy from experimental ice nucleation
data used by the authors suffers from significantly larger uncertainties than
reported in the text, rendering the method quite inaccurate for the estimation
of interfacial energy from experimental ice nucleation data.

Sincerely yours,

Tom\'a\v{s} N\v{e}mec

\subsection{Editor's decision -- July 11, 2013}
Dear Dr Nemec:

Manuscript ID CP-CMT-05-2013-052163
Title: Comment on "Kinetics of the homogeneous freezing of water" [Phys. Chem. Chem. Phys., 2010, 12, 10380-10387]

Thank you for your recent submission to Physical Chemistry Chemical Physics.  We have now received the reviewers' reports on your manuscript, which are copied below.  After careful evaluation of the manuscript and reports, I regret to inform you that we do not find your current manuscript suitable for publication in Physical Chemistry Chemical Physics. Further details regarding the reason for this decision are given in the reports.

I am sorry not to have better news for you and I hope the outcome of this specific submission will not discourage you from the submission of future manuscripts.

Yours sincerely,

M.S., Publishing Editor

\subsection{Referee 1 comments -- July 11, 2013} \label{ap:pccpr1}

This paper is incorrect and should not be published. The analysis outlined by the author is extremely naive and inappropriate, ignoring any literature regarding what it means to do a "fit" to experimental data. If the author feels that Murray et al.'s analysis is inappropriate, he should refit the data within the framework of linear (or non-linear) regression, accounting for the uncertainties in the experimental data, and explain why the approach used by Murray et al. in not good enough. Reference to the statistical literature is a must.

\subsection{Referee 2 comments -- July 11, 2013} \label{ap:pccpr2}

The author in his comment first argues that "two ice nucleation data sets that do overlap within their uncertainty intervals" lead to estimates of the interfacial energy that "do not overlap in their uncertainty intervals".  I do not see any contradiction here because the data sets are not identical, they have different 2D distributions and especially different ln J vs. T slopes (the slope m with respect to T\^{}-3 (ln S)\^{}-2 is proportional to the former slope).

Second, the author tries to estimate the uncertainties by drawing a parallelogram around the data set.  This method does not take into account statistical nature of the data and leads (for many data points) to  overestimated values.  The standard linear regression, as in the Murray et al. paper, is more appropriate.  The result m=-(6.02 +- 0.36) x 10\^{}7 K\^{}3 just means that the value of m is between [6.02 - 0.36, 6.02 + 0.36] with the probability of about 68 \%.

I therefore do not consider the comments relevant and do not recommend the Comment for publication.

According to my opinion, The Murray et al. method is not without possible pitfalls either.  A small systematic error in T (+- 0.6) does not affect the ln J vs. T slope nor the final interfacial energy value.  On the other hand, if also the x-data (i.e., T or T\^{}-3 (ln S)\^{}-2) were subject to statistical errors, the calculated $\left| m \right|$ would be underestimated.  The statistical part of the uncertainty in T is not discussed in the Murray et al. paper, and from Fig. 5 only an upper (likely very pessimistic) bound of about 0.2 K can be obtained; model data in the range of T of 1.8 and standard deviation of 0.2 lead to $\left| m \right|$  lower by 10-15 \% (depending on their unperturbed distribution).  In addition, I am concerned with the accuracy of the classical nucleation theory.

\subsection{Resubmission -- October 1, 2013}

Dear PCCP Editors,

I would like to submit the paper Comment on "Kinetics of the homogeneous freezing of water"
[Phys. Chem. Chem. Phys., 2010, 12, 10380--10387] by Tom\'a\v{s} N\v{e}mec
for your consideration for publication in PCCP as a comment.

This manuscript is a modification of the manuscript CP-CMT-05-2013-052163
that was rejected by PCCP in July 2013. In the modified manuscript I am carefully addressing the comments of both reviewers that criticized the statistical
analysis presented in the previous manuscript.

However, it was not the statistical analysis that was the main point of my
comment. The main point was the contradiction between the input and output
values of the CNT analysis presented by Murray et al. Regrettably, Reviewer
1 did not recognize this fact at all, and Reviewer 2 stated "I do not see any
contradiction here because the data sets are not identical, they have different
2D distributions and especially different ln J vs. T slopes". This statement does
not prove my argument wrong though. The CNT relation between the interfacial energy and nucleation rate is a bijective function and therefore any two
intersecting data sets necessarily have to be projected in two, also intersecting
data sets. This is generally valid regardless of the underlying statistical analysis
of the nucleation data, their distribution, the slope ln J vs. T, or the fact that
the two data sets are not identical. Therefore, I am also trying to elucidate the
contradiction in interfacial energy estimate more carefully in this new version
of my manuscript.

I do believe that my comment is addressing a flaw in the ice-water interfacial
energy estimation method of Murray et al., and since the method of Murray et
al. was already used by Manka et al. I feel it should be pointed out to the
nucleation community. Therefore I am asking for this second consideration of
my manuscript.

\subsection{Editor's decision -- November 20, 2013}
Dear Dr Nemec:

Manuscript ID CP-CMT-10-2013-054150
Title: Comment on "Kinetics of the homogeneous freezing of water" [Phys. Chem. Chem. Phys., 2010, 12, 10380-10387]

Thank you for your recent submission to Physical Chemistry Chemical Physics.  We have now received the reviewers' reports on your manuscript, which are copied below.  After careful evaluation of the manuscript and reports, I regret to inform you that we do not find your current manuscript suitable for publication in Physical Chemistry Chemical Physics. Further details regarding the reason for this decision are given in the reports.

I am sorry not to have better news for you and I hope the outcome of this specific submission will not discourage you from the submission of future manuscripts.

Yours sincerely,

T.S., Publishing Editor

\subsection{Referee 1 comments -- November 20, 2013} \label{ap:pccpr3}

Nemec reanalyses a homogenous nucleation data set published in PCCP by Murray et al. in 2010. Those authors report homogeneous nucleation rate coefficients determined from droplet freezing experiments. Murray et al. also analysed their own and some literature data in order to derive interfacial energies based on classical nucleation theory.  Nemec argues that they should have used a more rigorous statistical analysis which takes into account the uncertainty in temperature as well as uncertainty in nucleation rate. They produced a revised interfacial energy of 22.9 $\pm$ 1.3 mJ m-2 as opposed to the value of 20.8 $\pm$ 1.2 mJ m-2 determined by Murray et al.

In short, I do not think this paper reports an important new physical insight.

My key concern is that the revised value of the interfacial energy is within the estimated error as that reported by Murray et al.. i.e. Nemec's analysis seems to support Murray's result and show that their simpler statistical analysis produced a reasonable result. Given the scatter in the experiment and the experimental limitations imposed by the broad range of droplet sizes used by Murray et al. it seems an over interpretation to write a comment around such a minor shift in interfacial energy.

I would also like to caution against over-interpreting values such as the interfacial energy when deriveng it based on classical theory.  In order to do this one needs to make some rather crude assumptions.  It is necessary to assume a particular phase forms and takes on a particular shape and then use macroscopic values for its thermodynamic stability in order to derive the interfacial energy.  The validity of these assumptions needs to be borne in mind when interpreting the values of interfacial energy which emerge.  Given new information on the structure of cubic ice published since 2010 the assumption that the ice which nucleated in Murray's experiments is the same as that which Shilling et al. made (which was used to establish the saturation) now looks questionable. Given the experimental uncertainties and assumptions made in the use of classical theory it is an over-interpretation of these data to say that the value of interfacial energy should be shifted by 2.1 mJ m-2.

Minor points

Nemec states that S is the supersaturation, it is in fact the saturation. S-1 is the supersaturation.

\subsection{Referee 2 comments -- November 20, 2013} \label{ap:pccpr4}

This comment takes a narrow view of the subject it addresses. In experiments as difficult as those he takes under examination do not produce data points that can all be expected to have the same error limits over the whole range of the measurements. Thus, there is little point in arguing which type of regression analysis is to be used and that two data sets obtained with different methods have a minor disagreement. If the author intends to propose that a combined analysis of the two at a sets leeds to a value for the interfacial energy that is more reliable than values derive from either set alone, he should propose that without implying errors in the sources of the empirical data. He should then also defend the new value as being more in accord with other results, or with some theoretical argument, if he knowns of such.  As it is the Comment would contribute little to a productive discussion of the issues.

\subsection{Author's appeal -- December 1, 2013}

Dear PCCP editor,

on November 20, 2013, I received your decision rejecting the publication of my manuscript CP-CMT-10-2013-054150 titled Comment on "Kinetics of the homogeneous freezing of water" [Phys. Chem. Chem. Phys., 2010, 12, 10380-10387]. While analyzing the comments of the two referees, however, I came to the conclusion that your decision was based on comments of very low scientific value. I did not find a single argument disproving my results in the comments of the referees, and therefore I appeal against your decision and I request a new consideration of my manuscript by PCCP.

My main concern is that none of the referees has recognized the logical discrepancy I identified in the results of Murray et al., and I would like to note at this point, that I specifically mentioned this fact in the Cover Letter to my manuscript, I quote "The main point (of my comment) was the contradiction between the input and output values of the CNT analysis presented by Murray et al. ...".

Instead, Referee 1 is building up his arguments around his personal feeling that the new estimate of interfacial energy presents a minor shift; without realizing that the actual extent of the shift is irrelevant. Further, Referee 1 is mentioning his doubts about the assumptions of the classical nucleation theory and about the applicability of Shilling et al. experimental results, which are irrelevant since I use the same classical nucleation theory formalism and Shilling's parametrization as was used in the work of Murray et al. I comment on . Finally, Referee's 1 criticism of my definition of supersaturation is invalid.

Referee 2 is misinterpreting my arguments, he creates false presumptions that are not written in my comment. He criticizes me for not defending the new value as being more in accord with other results, while I actually do present the accordance of the new value with another theoretical result. This is unacceptable.

My detailed remarks to the comments of the two referees follow; the original comments of the referees being printed in italics.

Referee: 1

\textit{Nemec reanalyses a homogenous nucleation data set published in PCCP by Murray et al. in 2010. Those authors report homogeneous nucleation rate coefficients determined from droplet freezing experiments. Murray et al. also analysed their own and some literature data in order to derive interfacial energies based on classical nucleation theory.  Nemec argues that they should have used a more rigorous statistical analysis which takes into account the uncertainty in temperature as well as uncertainty in nucleation rate. They produced a revised interfacial energy of 22.9 $\pm$ 1.3 mJ m-2 as opposed to the value of 20.8 $\pm$ 1.2 mJ m-2 determined by Murray et al. }

This summary by Referee 1 ignores the main point, i.e. the logical contradiction in the Murray et al. analysis between input data (i.e. measured nucleation rates of Murray et al. and Stan et al.) and output values (i.e. estimated interfacial energies).

\emph{In short, I do not think this paper reports an important new physical insight.}

My paper is a comment. The comment addresses a logical contradiction found in the work it comments on, a contradiction that results in an inconsistent estimates of ice-water interfacial energy. Further, it shows how to avoid such inconsistencies.

\emph{My key concern is that the revised value of the interfacial energy is within the estimated error as that reported by Murray et al.. i.e. Nemec's analysis seems to support Murray's result and show that their simpler statistical analysis produced a reasonable result. }

The key concern of Referee 1 is just a statement of the obvious. The standard-deviation intervals of the  "revised value" and the value reported by Murray are indeed overlapping, i.e. 22.9 $\pm$ 1.3 mJ/m2 vs. 20.8 $\pm$ 1.2 mJ/m2. However, the discrepancy discussed in my comment lies in the standard-deviation intervals of interfacial energy estimate of Murray based on their own nucleation data, and based on the nucleation data of Stan, i.e.  20.8 $\pm$ 1.2 mJ/m2 vs.  23.7 $\pm$ 1.1 mJ/m2., which are indeed disjunct. And the logical contradiction lies in the fact that the experimental nucleation-rate data sets  used to deduce the two disjunct intervals are not disjunct, while at the same time the nucleation rate is a bijective function of interfacial energy.

\emph{Given the scatter in the experiment and the experimental limitations imposed by the broad range of droplet sizes used by Murray et al. it seems an over interpretation to write a comment around such a minor shift in interfacial energy.}

Based on my previous remark, the extent of the shift discussed by Referee 1 is irrelevant. However, even though the shift may seem minor in this case, in the case of Manka et al. (Ref.[5]) the discrepancies are larger. Manka et al. arrived with the method of Murray at the value 15.6 mJ/m2 while a theoretical method presented in Ref.[4], that is in my comment found to be consistent with the "revised value", gives an interfacial energy of 22.4 mJ/m2. This increase of 44\% can hardly be called a minor shift.

\emph{I would also like to caution against over-interpreting values such as the interfacial energy when deriveng it based on classical theory.  In order to do this one needs to make some rather crude assumptions.  It is necessary to assume a particular phase forms and takes on a particular shape and then use macroscopic values for its thermodynamic stability in order to derive the interfacial energy.  The validity of these assumptions needs to be borne in mind when interpreting the values of interfacial energy which emerge.}

The above-mentioned limits of classical nucleation theory (CNT) are generally known. It is not the intention of my comment to advance the theory of nucleation. The same CNT formalism is used in my comment as was used by Murray et al., and therefore the personal opinion of Referee 1 regarding CNT is irrelevant.

\emph{Given new information on the structure of cubic ice published since 2010 the assumption that the ice which nucleated in Murray's experiments is the same as that which Shilling et al. made (which was used to establish the saturation) now looks questionable.}

I use the same saturation pressure parametrization that was used in the work of Murray. Unless the new information about the structure of cubic ice is transformed into a more precise estimation of its saturation pressure than the result of Shilling, there is nothing questionable in this regard.

\emph{Given the experimental uncertainties and assumptions made in the use of classical theory it is an over-interpretation of these data to say that the value of interfacial energy should be shifted by 2.1 mJ m-2.}

Again, this conclusion of Referee 1 completely ignores the main point of my comment, i.e. the logical contradiction in the Murray et al. analysis between input data (i.e. measured nucleation rates of Murray et al. and Stan et al.) and output values (i.e. estimated interfacial energies).

Minor points

\emph{Nemec states that S is the supersaturation, it is in fact the saturation. S-1 is the supersaturation.}

I use the symbol S for supersaturation, i.e. the ratio of the actual vapor pressure p and the saturation pressure $p_{sat}$ (the ice-vapor equilibrium pressure in this case to be precise). This is the standard definition of "supersaturation" used throughout the nucleation literature. The term "saturation" refers to the thermodynamic state at the saturation line, i.e. when $p = p_{sat}$. Calling $S=p/p_{sat}$  "saturation", as the Referee 1 is suggesting, therefore makes no sense; doing so would make S always equal to 1.

Referee: 2

\emph{This comment takes a narrow view of the subject it addresses. In experiments as difficult as those he takes under examination do not produce data points that can all be expected to have the same error limits over the whole range of the measurements. }

This statement is manipulative. I am in no way assuming that the nucleation experiments produce data points "that can all be expected to have the same error limits over the whole range of the measurements". And I am definitely not making any conclusions based on such an assumption. I am working with the same error estimates as Murray et al. in their paper I am commenting on.

\emph{Thus, there is little point in arguing which type of regression analysis is to be used and that two data sets obtained with different methods have a minor disagreement.}

I am not making a point by arguing that "that two data sets obtained with different methods have a minor disagreement". I address an inconsistency in the method of Murray et al., who by employing a certain regression method arrive at contradictory values of interfacial energy.

\emph{If the author intends to propose that a combined analysis of the two at a sets leeds to a value for the interfacial energy that is more reliable than values derive from either set alone, he should propose that without implying errors in the sources of the empirical data.}

This paragraph is again manipulative; Referee 2 speculates about my intentions and accuses me of interpretations that are not written in my comment. I do not "propose that a combined analysis of the two data sets" leads to a more reliable interfacial energy estimate. I am definitely not "implying errors in the sources of the empirical data"; I work with error estimates reported by the respective experimentalists.

\emph{He should then also defend the new value as being more in accord with other results, or with some theoretical argument, if he knowns of such.}

Pointing out the accordance of the new value with other result is exactly what I am doing at the end of the third last paragraph. I quote: "The new ODR based estimate of ice-water interfacial energy is also consistent with my recent result [4] where the ice-water interfacial energy was estimated without the need to evaluate the slope of the experimental ice nucleation data." This argument of Referee 2 is completely flipped over.

\emph{As it is the Comment would contribute little to a productive discussion of the issues.}

My comment addresses a crucial contradiction in the analysis of the ice nucleation data for the estimation of ice-water interfacial energy which was used already by two research teams. Moreover, it gives a hint how to avoid reporting inconsistent results by using a data regression method compatible with the nature of the ice nucleation experimental data.

\subsection{Decision -- January 23, 2014}
Dear Dr Nemec:

Manuscript ID CP-CMT-10-2013-054150.R1
Title: APPEAL: Comment on "Kinetics of the homogeneous freezing of water" [Phys. Chem. Chem. Phys., 2010, 12, 10380-10387]

Thank you for submitting a revised version of your manuscript to Physical Chemistry Chemical Physics.  We have now received the appeal referee's report on your manuscript, which is copied below.  After careful evaluation of the manuscript and reports, I regret to inform you that we do not find your current manuscript suitable for publication in Physical Chemistry Chemical Physics. Further details regarding the reason for this decision are given in the report.

I am sorry not to have better news for you and I hope the outcome of this specific submission will not discourage you from the submission of future manuscripts.

Yours sincerely,

T.S., Publishing Editor

\subsection{Reviewer 3 comments -- January 23, 2014} \label{ap:pccpr5}

The single point raised by the author reduces to a disagreement between two values of a
parameter important in classical nucleation theory (CNT), namely the interfacial energy between
supercooled water and ice at low temperatures of interest (220 to 273 K). This becomes clear
once you filter the proposed comment from the notion "contradiction" and mathematical jargon
(e.g. "bijective" function which just expresses a 1:1 correspondence between the dependent and
independent variable in a simple monotonous function or "intersecting data sets" prohibited by
"logic"). There is no contradiction, what remains is only a disagreement between two parameter values. Natural science is deductive, not inductive: truth is gained from observations, not from speculative thinking. I fully disagree with the author that the article is addressing a flaw in the
ice-water interfacial energy estimation method of Murray et al. It is debatable whether or not to
use the (iteractive) ODR regression technique compared to the normal one, agreed. However, in
no case should a "contradiction" be construed on the base of two competing results (Murray et al.
vs. Stan et al.). The notion of contradiction conceals the truth about the scientific method based
on observation and approaching the true value within uncertainty limits.

The author fails to address the much larger problem of systematic error that has plagued the
measurement of ice nucleation rates over the years, namely the thermodynamic stability and state
of the newly formed solid phase and its slow conversion from ice I c to ice I h . This may certainly
influence the numerical value of the interfacial energy resulting from experiments to a much
larger extent than anything else. (See for instance the recent comprehensive and authoritative
publication by W. Kuhs and colleagues: PNAS 2012, 109, 21259-21264). As a chemical
kineticist I question the wisdom to extract a thermodynamic function from a T-dependence of just
1.7 (Murray et al.) or 2.2 K (Stan et al.) to any degree of accuracy, notwithstanding the fact that
the nucleation rate changes by 2 and 3.5 decades, respectively. Experimental measurements are
difficult and the extracted parameters such as the interfacial energy may be influenced by
systematic "errors" or uncertainties to a significant extent. There is no guarantee that the
resulting "ice" is identical in the bubble (Stan et al.) vs. the droplet experiments (Murray et al.).
Based on this the agreement between both data sets is remarkable, even without ODR applied to
the data of Murray et al.!

I fully concur with the opinion of both referees and believe that the revised version of the
referenced comment is unsuitable for publication in Physical Chemistry Chemical Physics.
There is too little substance for communication to the general PCCP readership. I would advise
the author to publish this comment to a specialized journal in the field.

In the end two minor comments:

- (Super)Saturation ratio ($p/p_{sat}$ ) and supersaturation are two different parameters.
"Supersaturation" is effectively $p/p_{sat} - 1$ (see e.g. Cloud Physics, Rogers and Yau, pg.
88). Therefore, referee 1 is correct, and the author seems to have fallen victim to sloppy
nomenclature.

- I am astonished about the author's arrogant tone of the rebuttal to the criticism of both
referees. I am not used to this and hope it will be left to this one-time derailment.

\section{Journal of Crystal Growth (October 2014 -- March 2015)} \label{ap:jcg}

\subsection{Manuscript submission -- October 23, 2014}
Dear Editor,

the submitted manuscript addresses an issue in evaluating interfacial energies from ice nucleation data that I discovered in published works. It points out the importance of a proper statistical processing of experimental nucleation data to arrive at a consistent estimate of the interfacial energy.

The paper was originally prepared as a short comment to the work Murray et al. [Phys. Chem. Chem. Phys., 2010, 12, 10380--10387] and submitted for publication in Phys. Chem. Chem. Phys. The publication was rejected by the PCCP referees. Since the comment was limited to two pages in length, I prepared this extended version addressing the main concerns of the PCCP referees and discussing the subject in more detail.

I took the liberty to include a mathematical proof in the text of the paper. Although the proof is a trivial exercise utilizing the properties of a bijective function, I am convinced it is valuable part of the paper that clearly demonstrates the logical contradiction in the results published by Murray et al.

\subsection{Editor's decision -- March 12, 2015}
Dear Dr. Tom\'a\v{s} N\v{e}mec,

Reviewers' comment on your work has now been received.  You will see that they are advising against publication of your work. I also agree with the opinion of the reviewer and I think that this paper should be submitted to PCCP journal, because the data previously published in PCCP were reanalyzed in this paper. Therefore I must reject it.

For your guidance, I append the reviewers' comments below.

Thank you for giving us the opportunity to consider your work.

Yours sincerely,

Y.F., Editor

\subsection{Reviewer's comments -- March 12, 2015} \label{ap:jcgr1}

Nemec presents a study of the derivation of interfacial energy from literature homogeneous nucleation data.  They conclude that using their revised analysis that the interfacial energy is 22.9 $\pm$ 1.3 mJ m-2 rather than the value of 20.8 $\pm$ 1.2 mJ m-2 reported previously for the same data set from Murray et al. (PCCP, 2010).  The key point that Nemec seems to miss is that these two values are the same within uncertainty.  There is some limited value in pointing out that data with inherent uncertainty can be analysed with different fitting routines to yield different slopes, but Nemec's values are within quoted uncertainties of the original.  Therefore this document does not reach the standard of a scientific paper, or even a technical comment, because it does not report anything new, it's main conclusion is wrong (in that Nemec claims a difference in values when they are in fact the same within error) and it is misleading in the claim that there is a problem with
the 'theoretical development of the method' (there is nothing wrong with the method). Hence, I  do not recommend it for publication.

Besides these fundamental floors, there are a number of specific issues which also preclude publication:

1)	The tone of the paper is not of a balanced scientific discussion, it comes across that Nemec has an axe to grind.  He seems to have taken the stance that much of the rest of the community are idiots and has then gone on to single out a few groups for attack.

2)	An example of aggressive attack is in the discussion of the proof of a 'logical contradiction' in the analysis of Murray et al.  Murray's paper has been singled out in the abstract, introduction, section 3 and the conclusions.  I read the section 'Proof of inconsistence' with great interest, waiting to find out what the fatal floor in logic was. It seems that the floor is the choice of fitting routine. This is not a floor in the 'slope method'. The language used by Nemec implies that there is a fundamental logical floor in the analysis, such as an assumption that cannot be justified or an error in the algebra.  At most the choice of fitting routine is a detail which should be discussed. The fundamental problem with the data of Murray et al. is there is a high degree of uncertainty and there will always be ambiguity and uncertainty in a data set where there is a lot of scatter.  Nemec is trying to over-interpret this data.  The fact that the revised analysis yields the
same values (within uncertainty) as the old analysis also shows that there is nothing of great importance to say.

3)	I took the opportunity to read over Murray's paper and I think that Nemec has missed or purposely ignored a few other facts.

a.	Murray et al. state that their T uncertainty is 0.6 K, not 0.4 K - does the use of an incorrect value impact the analysis of Nemec? Would the revised uncertainty on the revised interfacial energy be larger?

b.	Temperature uncertainty is discussed by Murray et al. as a major source of uncertainty: 'Future studies should focus on reducing uncertainty in temperature and extending reliable nucleation rate measurements over a wider temperature range'.

c.	Looking at Fig 4 of Murray et al., it seems that their data suffers from run to run variability.  This may be run to run variability in temperature offset.  Nemec is assuming that the temperature uncertainties are random.  They are not, the precision within a single run seems to be better than the accuracy. This leads to a complex situation and I am not sure which type of slope fitting would be most appropriate, but since the two fitting procedures produce the same values within uncertainty, then we have to conclude that the choice of fitting routine is of secondary importance (and certainly not the subject of an entire paper!).

d.	I have a criticism of Murray et al.'s data: they have used broad droplet size bins to work out J.  This is far from idea and will have the effect of over-predicting J at high temperatures and under-predicting at low temperatures.  The choice of fitting procedure is likely secondary to this.

4)	On p4 it is discussed that the fitting procedure results in a larger scatter in the prefactor for the literature data.  This statement is not substantiated. In order to show this, Nemec needs to reanalyse the literature data.

\subsection{Author's appeal -- March 17, 2015}

Dear editor,

thank you for the consideration of my work for publication in the Journal of Crystal Growth. However, I cannot accept the judgement of the reviewer. Please, allow me to react to the comments of the reviewer and to defend my work.

First, I would like to stress that each of the three statements that the reviewer uses as the basis to proclame that my work, I quote, \emph{"does not reach the standard of a scientific paper, or even a technical comment"}, is false:

1) \emph{"it does not report anything new"} -- My paper presents a thorough analysis of an existing method to deduce interfacial energies from the measured ice nucleation data, and reveals a contradiction in the results of this method, hinting to an internal inconsistency within the theoretical development of the method. A solid mathematical proof is given that the results of the method are contradictory. It is this proof of inconsistency, that is the main result of the work, not the use of some different fitting procedure, as the reviewer is claiming.

2) \emph{"it's main conclusion is wrong (in that Nemec claims a difference in values when they are in fact the same within error)"} -- It is really not the main conclusion of my work, that the new value of the interfacial energy estimated by me differs from the original value by Murray in any way. Although I do state in the conclusion that my new estimate is larger by roughly 10 \%, the extent of the difference is actually irrelevant. Yes, the new value and the old value are the same within error, as the reviewer  discovered, but the main conclusion concerns a different problem. The problem lies in the fact that the slope method estimate of the interfacial energy, based on Murray's nucleation data, and the estimate based on the nucleation data of Stan, are different, and do not agree within their uncertainties, while both data sets, i.e. Murray's and Stan's, are basically the same. In other words, you cannot deduce two distinct results (estimates of the interfacial energy) from two basically equal values (Murray and Stan data) according to some procedure (the slope method). Doing so makes the procedure inconsistent. This is the core of the contradiction, and all this is described in the paper in a proper mathematical language, i.e. in terms of the proof presented in section 2. The reviewer has missed completely the idea of the proof (he does not even mention the interfacial energy estimate based on the Stan data in his comments, and for some reason speaks only about Murray's paper). The fact that he claims that the difference in my new and Murray's old value of interfacial energy is the main conclusion of my work is his misinterpretation of my results. This misinterpretation is further iterated in reviewer's comments 2) and 3.c).

3) \emph{"it is misleading in the claim that there is a problem with the 'theoretical development of the method' (there is nothing wrong with the method)"} -- There are indeed problems in the theoretical development of the method as thoroughly described in the paper. The main problem lies in the freedom of the user to choose the fitting procedure, which introduces additional uncertainty in the estimation of the interfacial energy, not reflected in the uncertainty estimate reported by Murray (and therefore arriving at contradictory results). Further problems of the method are discussed in section 4.2, please, see the paragraphs starting "Not only the already discussed regression issue is the source of uncertainty in the slope method".

Second, it is extremely unfair by the reviewer to make his arguments around his feelings about the \emph{"tone of the paper"}. Are his scientific arguments alone not strong enough to prove my results wrong? What the reviewer labels as "an axe to grind" is, in my point of view, an effort to effectively communicate my results based on rational thinking and logic. Also, I definitely have not \emph{"singled out a few groups for attack"}. I am analyzing a certain method for the estimation of the ice-water interfacial energy that was published in the literature, and that was used in quite a few works, i.e. references [2,5,6,7,8,9], as summarized in the introduction. Since the method was used several times already, and I discovered a crucial issue (and I present ways how to avoid this issue), I think it is desirable for the community to learn the results.

Now, I would like to go back to your decision and the recommendation to publish my work in PCCP. I already tried to publish my results as a comment in PCCP as I brought to your attention in my cover letter. My paper was rejected as "unsuitable for publication in PCCP", and this is therefore not an option for me. What troubles me is that I got similar reviews from PCCP as the review you based your decision on. The similarity lies in that the review was always a mixture of misinterpretation of my results (my new value is not that different from the old one) and personal invectives (my arrogant tone). I have therefore a strong feeling that your reviewer is biased and already aware of my history in regard to PCCP.  I think this gives me the right to ask you for an additional review by a person that will judge my results from an independent scientific perspective and focus on confirming/disproving the contradictions I discovered in the results of the slope method.

Thank you for you time and looking forward to your answer,

Tomas Nemec.

\subsection{Editor's reply}
No reply.

\section{Advances in Physical Chemistry (July 2016 -- June 2017)} \label{ap:apc}

\subsection{Manuscript submission -- July 11, 2016}
Dear Editor of APC,

I would like to submit my work titled "Critical analysis of the slope method for estimation of ice-water interfacial energy from ice nucleation experimental data" for your consideration to be published in APC. The paper addresses an issue in evaluating interfacial energies from ice nucleation data that I discovered in recent works. It points out the importance of a proper statistical processing of experimental nucleation data to arrive at a consistent estimate of the interfacial energy.

The paper was originally prepared as a comment to the work Murray et al. [Phys. Chem. Chem. Phys., 2010, 12, 10380--10387] and submitted for publication in Phys. Chem. Chem. Phys. The publication was rejected by the PCCP. Since the comment was limited to two pages in length, there was not enough space to discuss the issues of the slope method of Murray et al. I therefore prepared this extended version addressing the main concerns of the PCCP referees and discussing the subject in more detail.

Best regards,

Tomas Nemec

\subsection{Editor's decision -- June 8, 2017}
Dear Dr. N\v{e}mec,

After reviewing your Research Article 6498962 titled "Critical analysis of the slope method for estimation of ice-water interfacial energy from ice nucleation experimental data", I regret to inform you that it was found to be unsuitable for publication in Advances in Physical Chemistry. Please find attached the review report received.

Thank you for submitting your manuscript to Advances in Physical Chemistry.

Best regards,

N.E., Editorial Office

\subsection{Reviewer's comments -- June 8, 2017} \label{ap:apcr1}

N\v{e}mec presents a reanalysis of some homogeneous freezing data and reports a
revised interfacial energy for supercooled water-ice for one particular dataset. The
key point seems to be that Murray et al. (2010) used an ordinary least squared
routine whereas N\v{e}mec uses an orthogonal least squared routine to fit a straight line
to a dataset. N\v{e}mec's point that there may be better ways of fitting data such as this
is a valid argument, but there are some serious issues with the paper as it stands.
N\v{e}mec motivates his study by claiming Murray et al. produced values which are
'inconsistent', or that there is a 'logical contradiction' and then goes on to suggest a
different approach for fitting data. The fitting of the data and the revised interfacial
energy is somewhat useful and could be published, but the motivation needs to be
dramatically modified. In order to get this paper into a state where it could be
published N\v{e}mec needs to address these issues:

1. The claims throughout the paper that Murray et al. produced values which are
'inconsistent', or that there is a 'logical contradiction' is wrong. I read the
abstract and then the conclusions first and was concerned that N\v{e}mec's tone
implied Murray had made a serious error or mistake in his paper and then
read on to try to figure out what this major error was. I was surprised that the
mistake Murray et al. apparently made was simply choosing one method of
fitting over another method of fitting a dataset with a lot of scatter in it. It is
valid to propose a different fitting routine, but the orthogonal approach is not
unambiguously the correct approach to use and the the language needs to be
modified. N\v{e}mec must not state that there is a logical contradiction in Murray
et al. Citing Murray et al. in the abstract seems unnecessarily aggressive and
should be removed. The reference to an 'inconsistent estimate of the ice-
water interfacial energy' has to be removed (see next point).

2. It must be stated what has been assumed. N\v{e}mec has assumed that the
scatter in the data is entirely random. This is not the case. I have reproduced
Fig 4 from Murray et al 2010 below. This plot seperates the data into the 6
different experiments and the two size bins. What you see is that there is a lot
less scatter within one experiment and in one size bin (e.g. the blue triangles),
whereas there are systematic errors between experiments and between size
bins. A simple statement along the lines that N\v{e}mec makes the assumption
that the error is random, but in fact much of the spread was due to run-to-run
variability.

\begin{figure}[t]
	\includegraphics[width=\gs]{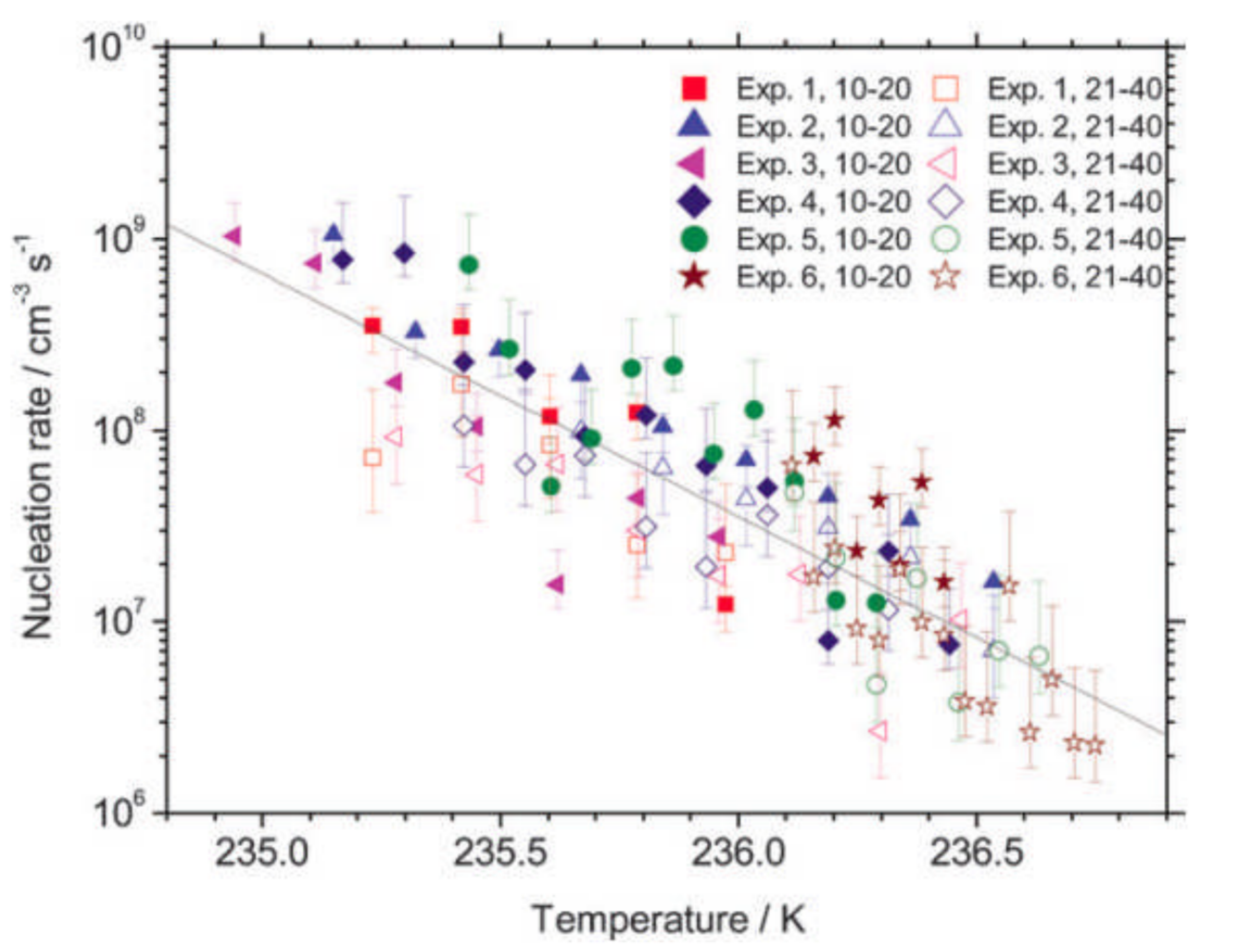}
\end{figure}

3. The 'proof' in section 3 which is presented should be deleted in its
entirety. This pedantic discussion boils down to the statement that $\gamma_M$ = 20.8
$\pm$ 1.2 mJ/m2 whereas $\gamma_s$ = 23.7 $\pm$ 1.1 mJ/m2. Yes, there should be a single
value, but these two values are not incompatible. My understanding is that
these errors are 1$\sigma$ (i.e. 68\% confidence interval), within 2$\sigma$ the values are
entirely consistent with one another. i.e. on the 95\% confidence interval the
two values are within uncertainty. This does not represent a 'logical
contradiction' or an 'inconsistency'.

4. The group of Murray recently published a new study of homogeneous freezing
with a new and improved methodology which has dramatically reduced the
scatter in the J values (Atkinson et al., 2016). These authors clearly discuss
the limitations of the old Murray et al. 2010 data. The new data is in very
good agreement with the nucleation data of Stan et al. Atkinson et al. discuss
(on p 6516) the Murray et al. 2010 data and suggest that the smaller slope in
Murray 2010 was due to issues with using size bins with broad droplet size
distribution and median droplet volumes. It is fair to say that the choice of
fitting routing could also have been a factor. Nevertheless, the key limitation of
the old Murray 2010 data is the large scatter in the data, which then means
the result is sensitive to the choice of fitting routine.

5. On p5 N\v{e}mec state that the scatter in J$_0$ is due to errors in the slope method.
This is just wrong. These estimates were made from the analysis of
numerous literature datasets. The uncertainty comes from uncertainty in the
data, due to this being a challenging measurement to make. If N\v{e}mec think it
is simply the fitting routine that is at fault he should refit all of the old literature
data and show this, rather than simply asserting it.

6. P5. Discussion of Manka data: N\v{e}mec has misinterpreted how Manka et al
derived their interfacial energy. Manka et al. did not use the lnJ vs T$^{-3}$ (lnS)$^{-2}$
analysis. When they refer to the 'same formalisms' as Murray et al., they
mean they used the same classical theory equations to describe J over a very wide range of temperatures. This whole paragraph needs to be deleted or
rewritten.

7. P6. A brief discussion on what may happen to other critical parameters in
nucleation theory with increasing pressure is needed. This was recently
discussed (Laksmono et al., 2015) and it is not obvious that one should try to
fit the low temperature nanodrop data to the same curve as the higher
temperature micron droplet data. This is controversial and the the group of
Manka et al. (led by Wyslouzil) have argued that pressure is not important for
the thermodynamic properties of water.

8. P2. When referring to the 'discussion' of the character of ice that nucleates,
N\v{e}mec needs to refer to more recent literature, such as the review of (Malkin
et al., 2015) on which Molinero was also a co-author and refers back to the
Moore and Molinero paper.

9. In the final concluding statements it is stated 'using a theoretical model of the
nucleation rate pre-factor instead of its fit from experimental nucleation data,
present a safer way of deducing the interfacial energy'. Koop and Murray
recently made an attempt in this direction, this should be mentioned (Koop
and Murray, 2016). They established the temperature dependence of all the
key terms in classical theory theoretically and then fitted the curve to the
available data. The only adjustable parameter was the interfacial energy at a
reference temperature.

On a more general note (and I do not wish to come across as patronising and
apologise in advance if I do), N\v{e}mec could make a more valuable contribution to the
field of ice nucleation if he were less aggressive in his writing. It is possible to get
ideas across without using language which could be interpreted as aggressive and
inflammatory. Referring to logical contradictions and inconsistencies in a particular
paper in order to justify this new approach is far too aggressive and there is no need
for it. A more neutral tone is always preferable in a scientific document. In this
particular paper, there was no need to try to imply that Murrray et al. (2010) was
wrong, but rather point out what you think is a better approach. i.e. the paper could
be written in a more positive way.

References

Atkinson, J. D., Murray, B. J., and O'Sullivan, D.: Rate of Homogenous Nucleation of
Ice in Supercooled Water, The Journal of Physical Chemistry A, 120, 6513-6520,
2016.

Koop, T. and Murray, B. J.: A physically constrained classical description of the
homogeneous nucleation of ice in water, J Chem. Phys., 145, 211915, 2016.

Laksmono, H., McQueen, T. A., Sellberg, J. A., Loh, N. D., Huang, C., Schlesinger,
D., Sierra, R. G., Hampton, C. Y., Nordlund, D., Beye, M., Martin, A. V., Barty, A.,
Seibert, M. M., Messerschmidt, M., Williams, G. J., Boutet, S., Amann-Winkel, K.,Loerting, T., Pettersson, L. G. M., Bogan, M. J., and Nilsson, A.: Anomalous
Behavior of the Homogeneous Ice Nucleation Rate in "No-Man's Land", J. Phys.
Chem. Lett., 6, 2826-2832, 2015.

Malkin, T. L., Murray, B. J., Salzmann, C. G., Molinero, V., Pickering, S. J., and
Whale, T. F.: Stacking disorder in ice I, Phys. Chem. Chem. Phys., 17, 60-76, 2015.

\end{document}